# Development of suspended core soft glass fibers for far-detuned parametric conversion

Anupamaa Rampur[1,2], Piotr Ciąćka[1], Jarosław Cimek[1,2], Rafał Kasztelanic[1], Ryszard Buczyński[1,2], Mariusz Klimczak[1,*]

[1]Institute of Electronic Materials Technology, Glass Department, Wólczyńska 133, 01-919 Warsaw, Poland
[2]University of Warsaw, Faculty of Physics, Pasteura 5, 02-093 Warsaw, Poland

*mariusz.klimczak@itme.edu.pl

**Abstract**

Light sources utilizing $\chi^{(2)}$ parametric conversion combine high brightness with attractive operation wavelengths in the near and mid-infrared. In optical fibers it is possible to use $\chi^{(3)}$ degenerate four-wave mixing in order the obtain signal-to-idler frequency detuning of over 100 THz. We report on a test series of nonlinear soft glass suspended core fibers intended for parametric conversion of 1000-1100 nm signal wavelengths available from an array of mature lasers into the near-to-mid-infrared range of 2700-3500 nm under pumping with an erbium sub-picosecond laser system. Presented discussion includes modelling of the fiber properties, details of their physical development and characterization, as well as experimental tests of parametric conversion.

## 1. Introduction

High brightness light sources covering long-wave near-infrared (NIR) and mid-infrared (MIR) wavelengths are attractive for applications in industry, military and sensing. Depending on particular technology, commercial devices in this group can be found operating at wavelengths between 2 and 20 µm, which is the molecular fingerprint region with numerous chemical specimens having their vibrational transitions in that part of the spectrum [1]. This is important for diverse settings, from real-time detection of explosives [2] to food quality control [3]. Another important applications of MIR light sources include, apart from general spectroscopy, agriculture [4], gas sensing [5], bio-medical research [6] and free space communication [7].

Light at the MIR wavelengths can be generated in trivalent rare earth-doped ions like $Nd^{3+}$, $Yb^{3+}$, $Er^{3+}$, $Tm^{3+}$ and $Ho^{3+}$ with proper choice of the host optical material. Specifically, it is obtained from direct transitions within the energy structure of the trivalent ions embedded in a host crystalline lattice [8]. Divalent optical materials, e.g. $Cr^{2+}$: ZnSe and $Fe^{2+}$: ZnS(e) are also attractive for MIR devices, opening operation wavelength ranges from 2 um to 3.5 µm and around 3.5 µm to 5 µm, respectively [9]. The quantum cascade lasers, where output is produced by consecutive radiative transitions within the conduction band, instead of the electron-hole recombination, are yet another alternative, with wavelength coverage between 3 µm to 25 µm [10].

High brightness MIR light can be obtained also with the use of nonlinear optical frequency conversion of pump and signal wavelengths. These would be typically available from mature light sources, i.e. high average power ultrafast lasers. Significant progress in this field has been obtained with non-centrosymmetric crystals having $\chi^{(2)}$ nonlinearity and exploiting difference frequency mixing, e.g. in periodically poled LiNbO$_3$ [11–13]. In optical fibers the $\chi^{(2)}$ nonlinearity cannot be used because glass is centrosymmetric. Nevertheless, optical fiber-based parametric amplification through the $\chi^{(3)}$ degenerate four-wave mixing (DFWM) has now been well recognized as means of efficient conversion in the NIR telecommunication wavelengths, enabling several to a dozen THz of detuning [14–16].

More recently, engineering of dispersion, nonlinearity and mode structure in photonic crystal fibers (PCF) resulted in theoretical reports on the possibility of far-detuned parametric conversion with signal-idler detuning in excess of 160 THz [17], followed by experiments demonstrating conversion either in the NIR or MIR wavelengths. Specifically, Nodop et al. reported on endlessly single mode silica photonic crystal fiber, in which an idler wavelength at over 2500 nm could be obtained under pumping with a standard wavelength of 1064 nm [18], while Godin et al. demonstrated conversion of a 2620 nm pump into over 3500 nm in a chalcogenide glass microwire developed from a step-index MIR fiber [19]. Parametric conversion in chalcogenide microwires was also reported under pumping at a standard wavelength of around 1900 nm from a mode-locked fiber laser, covering signal and idler wavelengths of around 1400 and 2900 nm, corresponding to a soft glass-fiber record, nearly 50 THz detuning range [20].

In this work, we investigate suspended core fibers made with lead-bismuth-gallium heavy metal oxide glass, for parametric conversion under pumping with a standard laser wavelength in the NIR range. The choice of glass material is motivated by its relatively high recrystallization resistance, ease of handling and good mechanical properties compared to chalcogenides, as well high nonlinearity and combination of visible to mid-infrared transmission, compared to silica glass. The fiber geometry involving large air-fill fraction has been suggested previously as adequate for parametric conversion due to particular dispersion shaping feasibility favoring DFWM [19]. The combination of the glass and the fiber type selected in this work further benefits from the potential to use a standard wavelength, i.e. 1560 nm available from maturing mode-locked fiber-based lasers, as the pump wavelength. Our discussion includes modelling of linear properties of different sets of geometric properties of the chosen fiber layout, in context of obtainable dispersion profile, as well as wavelength dependence of the effective mode area. Parametric conversion performance of selected fibers was then investigated with nonlinear propagation simulations. It is followed by characterization of the physically developed fibers and parametric conversion experiments.

## 2. Theory

DFWM in $\chi^{(3)}$ regime produces Stokes (idler) and anti-Stokes (signal) frequency sidebands of the component at the pump frequency. Without any intentionally introduced seeds, these sidebands are seeded by noise and the entire process is equivalent to inducing of modulation instability. When the pump wavelength falls into anomalous dispersion wavelengths of the nonlinear medium, such as an optical fiber, the signal and idler MI components are typically located closely around the pump wavelength [21]. The detuning between signal/pump and pump/idler components can be increased by pumping in the normal dispersion regime with an endlessly single mode fiber [18,22]. In an endlessly single mode fiber, an efficient signal and idler far away from the pump wavelength can be generated in which the effective mode field is independent of the wavelength, which further helps in efficient phase matching when pumped closer to zero dispersion wavelength (ZDW). This has been previously discussed in chalcogenide microwires and photonic crystal fibers [19,20,23]

Mathematically parametric conversion of the pump light by degenerate four-wave mixing in a $\chi^{(3)}$ nonlinear medium, such as a PCF, is described as follows:
$$2\,\omega_p = \omega_s \mp \omega_i \quad \#(1)$$
with $\omega_p$ being the angular frequency corresponding to the pump wavelength, and $\omega_s$, $\omega_i$ stand for angular frequencies of the signal and idler wavelength components. In order to discuss the phase matching condition for the DFWM, the detuning is introduced with the following formula:
$$\Omega = \omega_p - \omega_i = \omega_s - \omega_p \quad \#(2)$$
The phase matching condition is then defined by a dispersion relation obtained for a complex amplitude of a nonlinearly propagating pulse, perturbed at frequency $\Omega$ [19,24].

$$k = 2\gamma P_o(1 - f_r) + 2\left[\frac{\beta_2\Omega^2}{2} + \frac{\beta_4\Omega^4}{24}...\right] \quad \#(3)$$

where $\beta_{j's}$ are even dispersion parameters, $\gamma$ is nonlinear coefficient, and $P_o$ is the pulse peak power. Parametric wavelengths (anti-Stokes signal and Stokes idler) are obtained as a function of the pump wavelength using (2) and the detuning frequency is obtained from phase matching condition (3) for k=0. Further, for normal dispersion pumping, where $\beta_2$ is positive, from equation (3) it stems, that $\beta_4$ has to be negative, in order to obtain phase matching.

## 3. Fiber design

Far-detuned, NIR/MIR parametric conversion, demonstrated in a chalcogenide glass fiber micro-wire indicated the necessity for the fiber structure to feature large air-filling fraction [19]. Therefore, we adapted a fiber design based on the traditional suspended core geometry. The fiber was designed to have a central core, surrounded by 6 large air holes, as shown in a scanning electron microscopy (SEM) image in figure 1. In order to obtain large air fill fraction, the struts should be thin, hence we increased their number for a more reliable support of the core. Different structure geometries were modelled by varying core diameter and strut thickness. The strut thickness taken in simulations ranged from 0.2 to 1μm with a step of 0.2. Core diameters of 3, 4 and 5 μm were investigated. We use a labelling scheme SxdyRz to define these different fiber geometries. The x for 6 air holes is fixed, y stands for strut thickness and z for the core diameter and these values differ among the designs. Linear simulations were performed using Comsol/Matlab environment for a wavelength range from 500 to 5000 nm. The glass chosen for the simulations, and later for fiber development, was the lead-bismuth-gallium heavy metal oxide glass with chemical composition of $40SiO_2$, $30PbO$, $10Bi_2O_3$, $13Ga_2O_3$, $7CdO$. This soft glass combines good mechanical and rheological properties, moderate fiber drawing temperature of around 800°C, high nonlinear refractive index of $1.95\times10^{-19}$ $m^2/W$ measured at 1064 nm [25] and has been successfully applied in development of photonic crystal fibers for supercontinuum generation across the near-infrared [26,27]. The Sellmeier coefficients measured for this glass and used in the simulations were: B1=2.01188143, B2=0.54673236, B3=1.39488613, C1=0.01537572 $\mu m^2$, C2=0.06355233 $\mu m^2$, C3=141.6540462 $\mu m^2$. Calculated dispersion profiles shown in figure 2(a) for the core diameters of 3 μm, 4 μm, and 5 μm have a flattened "saddle" range of wavelengths between roughly 2500 nm and 4000 nm. This "saddle" area becomes more curved with decreasing core diameter.

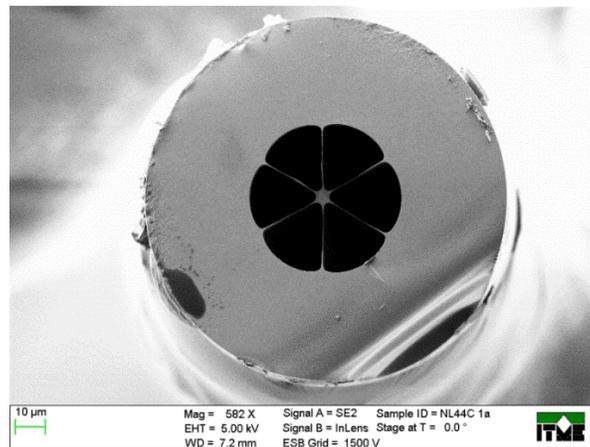

**Figure 1.** SEM image of a typical fiber structure developed based on the discussed designs. Fiber outer diameter in the image is 120 μm, the fiber core is 4.5 μm and strut thickness is about 81 nm.

The dispersion profiles reach a local maximum of the "saddle" section at different wavelengths, depending on the geometrical parameters, i.e. the core diameter and the strut thickness. Various possible combinations of these geometrical parameters have been studied using numerical simulations, in context of their influence on the dispersion characteristics, i.e. their ZDW location and the location of the local maximum in the "saddle" of the profile. Results of these simulations are shown in figure 2(b,c) respectively for the dependence of the dispersion local maximum and ZDW location for analyzed combinations of the fiber core diameter and the strut thickness. For an increase in strut thickness from 0.2 µm to 1.0 µm and for increasing in core diameter from 3 µm, through 4 µm to 5 µm, the dispersion maximum is decreasing rapidly. Increasing the core diameter results in red shifting of the ZDW, although its location remains practically unchanged across the investigated strut thicknesses for the largest fiber core. These trends are expected in a way, that thicker details of the fiber structure draw the dispersion profile closer to the material dispersion of the glass. In particular, the 5 µm core becomes large enough to contain the mode and the varying strut thickness of different hypothetical fibers has little effect either on the local maximum nor on the ZDW.

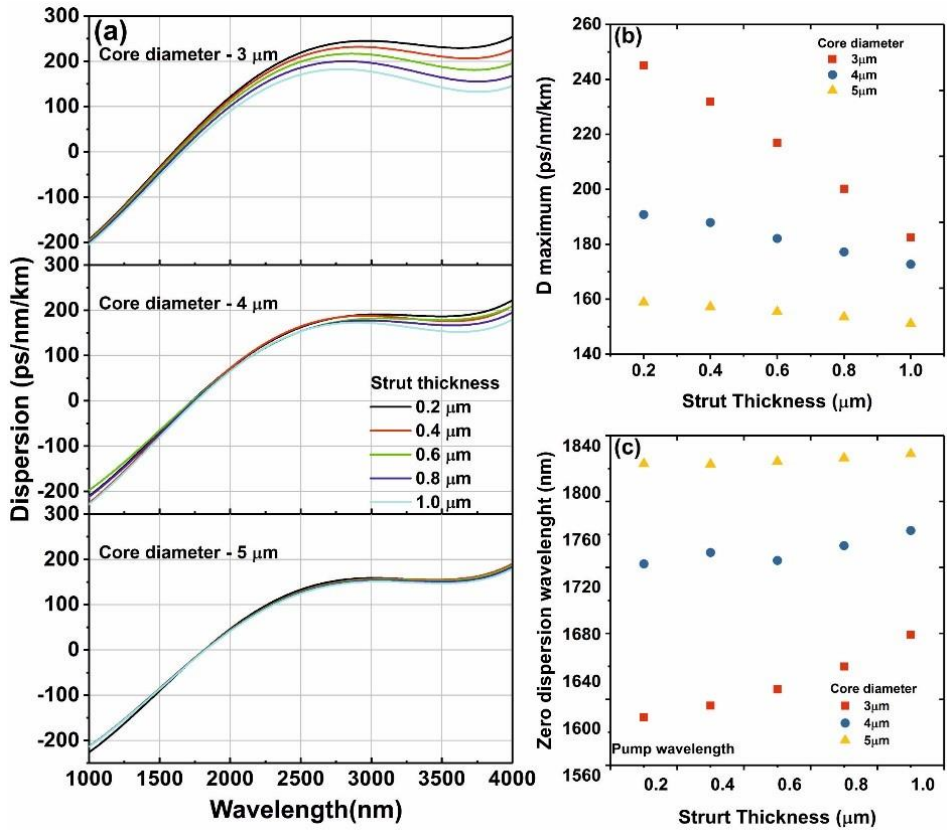

**Figure 2.** (a) Calculated dispersion profiles for different strut thickness from 0.2 µm to 1 µm and core diameter 3µm, 4µm, and 5µm. (b) Values of dispersion at the "saddle" maximum for different strut thicknesses. (c) Zero dispersion wavelength positions for different core diameters and strut thicknesses.

## 4. Phase matching and mode area matching in the designed fibers

Far-detuned parametric conversion in optical fiber-based structures has been shown most effective, when the ZDW of the structure's positive slope dispersion profile is redshifted by single tens of nm from the pump laser wavelength [17] The ZDWs of the designed fibers fall between 1600 nm and 1850 nm, which motivates use of 1560 nm pump wavelength, available from a multitude of ultrafast mode-locked lasers.

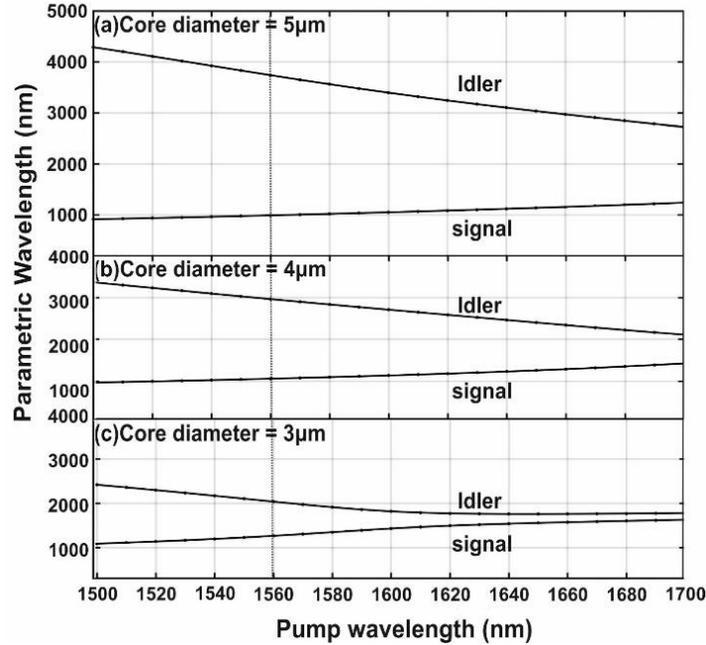

**Figure 3**: Phase matching curves for fiber having strut thickness: 0.2 μm and core diameter: a) 5 μm b) 4 μm c) 3 μm.

Phase-matching profiles were calculated for selected fiber structures using formula (3) are shown in figure 3. A fixed value of the nonlinear coefficient γ was assumed for the pump wavelength of 1560 nm, and was equal to 136 $km^{-1}W^{-1}$, 80.79 $km^{-1}W^{-1}$ and 53.3 $km^{-1}W^{-1}$ respectively for fibers with 3 μm, 4 μm and 5 μm core diameters. Obtained phase matching curves indicate theoretical possibility to obtain idler four-wave mixing signals in the mid-infrared (around 4000 nm), as well as in the near-infrared, around 2500 nm, depending on the core size and thus the ZDW. Large wavelength separation of the pump and idler signal results from increased wavelength separation between the pump wavelength and the ZDW, as is in the case of phase-matching profiles shown in figure 3 (a) 5 μm core, ZDW=1816 nm, (b) 4 μm core, ZDW=1731 nm and in (c) 3 μm core, ZDW=1600.7 nm. Efficient parametric conversion of a seeded signal component however requires not only phase matching, but also mode field overlap in order for the process to take place with any practical gain [18]. Suspended core fibers support high confinement of the guided mode, but a mode field mismatch between widely separated wavelengths, i.e. the pump and the idler must be considered. Figure 4 shows calculated effective mode areas for the entire series of the considered fibers. Dependence of the effective area on wavelength is important in a parametric converter fiber, because overlap between the signal and idler wavelengths, as well as between the signal/pump and pump/idler mode areas determines the efficiency of the conversion [18]. Suspended core fibers are characterized by strong confinement and thus all of the fibers are characterized with only moderate dependence of mode area on wavelength. As can be seen in figure 4, the effective area in the case of the 3 μm core fibers changes from between 1100 nm and 2600 nm from about 5.5 $μm^2$ to 6.5 $μm^2$. The two wavelengths correspond roughly to the maximally detuned FWM signal and idler wavelengths under 1560 nm pumping for the 3 μm core fibers, as shown in figure 6. The 3 μm core fiber series shows increased wavelength dependence for variants with thicker struts, which is explained by stronger mode penetration into the struts as the wavelength increases. In the case of fibers with core diameters of 4 μm or 5 μm this effect is much less pronounced. However, calculated phase matching wavelengths allow to expect larger detuning between signal and idler components, for example from 1100 nm to about 3500-4000 nm (figure 6). In such a case, the mode area mismatch would be larger than for the 3 μm core fibers, i.e. from around 14 $μm^2$ at about 1100 nm to 18 $μm^2$ at 4000 nm. We note that this would be unfavorable in a parametric amplification scenario, where for example a continuous-wave signal wavelength is in-coupled into the fiber, along with the pulsed pump, in order to amplify the red-shifted idler.

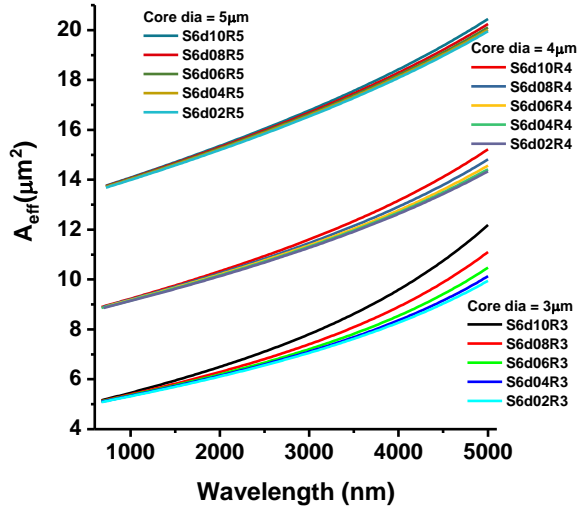

**Figure 4**. Calculated effective mode areas of investigated 6-strut suspended core.

Photonic crystal fibers with large air-fill fraction have been indicated as a good choice for $\chi^{(3)}$ parametric conversion because dispersion profiles for large detuning are readily obtainable in such structures [19]. The potential for high mode overlap between the three wavelength components, stemming from low wavelength dependence of the effective mode area, is also an obvious advantage. The mode structure of the designed fibers is also important, because excitation of a higher order mode of the fiber at the short wavelength component would result in poor mode overlap with the idler component. According to numerical simulations, the fibers designed in this work can support multiple higher order modes as shown in figure 5(a) for a 4 μm core fiber with 0.2 μm struts.

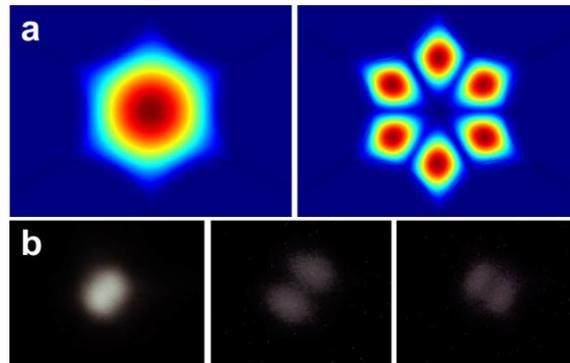

**Figure 5**. Mode profiles (a) calculated for one of the developed fibers at a wavelength of 1000 nm and (b) observed experimentally at a wavelength range of 1200-1300 nm using standard camera.

Experiments with physically developed fibers confirmed the multimode nature of these fibers, as shown in figure 5(b). The shown images were taken in a fiber pumped strong enough for supercontinuum generation to occur. It was possible then to obtain the images shown in figure 5(b) with a standard camera, a 1200 nm high pass filter was also used. This experiment also revealed, that selectively single-mode operation can be obtained in the fiber upon proper adjustment of the in-coupling condition. This was further supported with numerical aperture (NA) measurements. The measurements were performed by knife-edging the output beam from one of the fibers (labelled NL44C1a in a following section and also shown in SEM image in figure 1), whereas a 1314 nm signal from a single-mode diode laser was in-coupled into the fiber. The value of NA measured in such conditions was 0.29.

In order to attempt a general assessment of how the detuning of the idler signal from the pump wavelength should depend on the position of the local maximum of the dispersion profile, we analyzed the positions of the idler signals, depending on the local maximum of the fiber dispersion profile. A summary of the idler and signal wavelengths for the 1560 nm pump wavelength is shown in figure 6 for all the considered fiber geometries as a function of the local maximum dispersion value. We note, that these local maxima are located at slightly different wavelengths in the anomalous dispersion range redshifted from the ZDW. It can be noted, that the dispersion value and thus the total group delay introduced in the parametric process, has strong effect for smaller detuning, when the pump wavelength-ZDW separation is also small. Here, for the 3 µm core fiber with the most blue-shifted ZDWs among the considered fibers (ZDWs between 1600 nm and 1680 nm) the idler wavelength expected from the phase-matching condition (3) changes from about 2100 nm for a fiber with local maximum dispersion value of 240 ps/nm/km, up to about 2500 nm for the fiber with local maximum dispersion of 180 ps/nm/km. We note also, that the fiber with this local maximum dispersion value is the fiber with the thickest struts of 1 µm. This particular 3 µm core diameter fiber has therefore the largest overlap of the guided mode with the glass, and indeed the fiber's ZDW has the most blue-shifted ZDW at 1680 nm among the 3 µm core fibers.

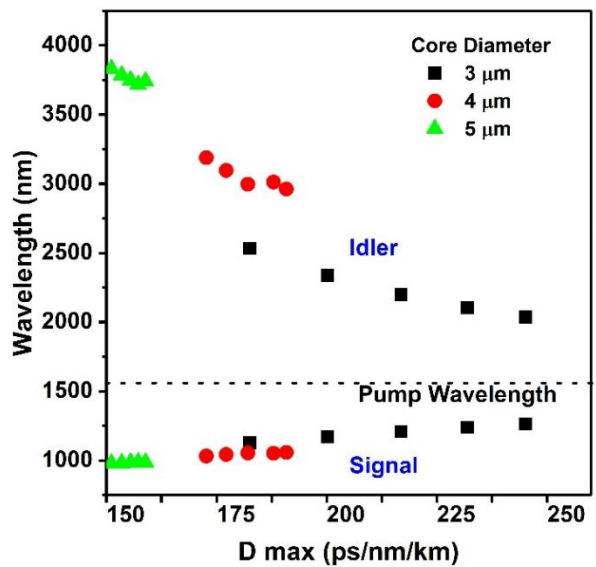

**Figure 6**. Signal and idler positions with respect to the D maximum.

## 5. Simulations of nonlinear frequency conversion

We used nonlinear propagation simulations in order to assess how different pump pulse durations and energies would shape performance of parametric conversion in the designed fibers. The modelling was performed using Generalized Nonlinear Schrödinger Equation (GNLSE) and the particular realization proposed by Travers, Frosz and Dudley was adapted [28]. In order to simulate inducing of modulation instability, the model was extended by inclusion of one-photon-per-mode noise. Furthermore, wavelength-dependent attenuation and effective mode area were included in the simulation to account for respectively the OH⁻ absorption line at around 2.8-3.0 µm and the effect of mode-overlap between the pump and idler wavelength components. The length of the fiber sample was selected at 15 cm due to the convenience of handling in a physical experiment. The relatively high nonlinear coefficient of the lead-bismuth-gallium oxide glass corresponding to a high nonlinear coefficient (136 km$^{-1}$W$^{-1}$ for the 3 µm core fiber with 0.2 μm struts) also result in very short nonlinear characteristic length scale of 0.45 cm, favoring a scenario in which the parametric conversion does take place over such short lengths of fiber.

The simulations were performed for pump pulse lasting 10 ps, to keep self-phase modulation at a disadvantage, and with the pump pulse energy varying from 1 nJ to 20 nJ. Depending on given dispersion profile of the fiber, a certain minimum pump pulse energy was necessary to obtain parametric conversion within a reasonable dynamic range of -20 dB relative to the pump intensity. It is to be noted, that $\chi^{(3)}$ based parametric conversion in chalcogenide microwires has been reported experimentally with the idler intensity roughly 40 dB below the pump intensity [19].

Figure 7 shows numerical simulation results for the fiber structure with a 4 μm diameter core and 1 μm thick struts. This particular set of parameters of the fiber geometry yields a dispersion profile with a ZDW located at 1760 nm and consequently the simulation allows for large idler detuning up to around 3200 nm. Because of large wavelength separation between the pump and the ZDW (roughly 200 nm), at ultrashort pump pulse conditions (close to 1 ps or shorter), a fiber having normal dispersion at the pump wavelength, positive dispersion slope and normal dispersion over 200 nm towards the red from the pump, should be expected to behave similarly to normal dispersion supercontinuum generation scenario, with self-phase modulation and optical wave-breaking distributing the input energy across wavelengths adjacent to the pump wavelength [29]. For this reason, the pump pulse duration assumed in the simulation was 10 ps. With 20 nJ of in-coupled energy, as shown in figure 7 (a), the redshifted idler component at 3200 nm has intensity of about 20 dB smaller than the 1560 nm pump signal. All the three components: the pump, the signal at around 1000 nm and the idler, are spectrally broadened around their central wavelengths due to modulation instability. Simulation performed for a 10 nJ input pulse revealed narrowband spectral features, however the idler intensity was about -80 dB below the pump intensity, as shown in figure 7 (a).

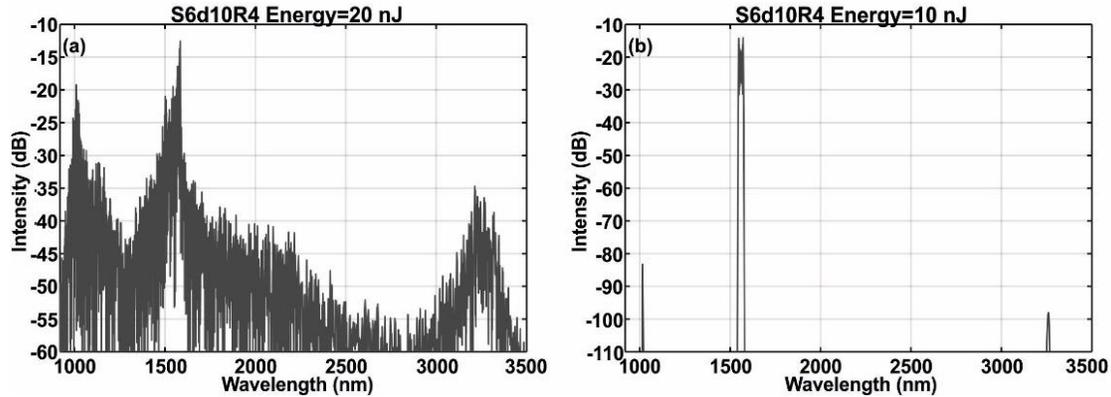

**Figure 7**. Simulated spectrum showing the parametric conversion in fiber having core diameter 4μm and strut width 1 μm, when pumped with 1560nm having pulse width of 10ps with energy a) 20 nJ and b) 10 nJ.

For the fiber geometry with 4 μm core and 0.8 μm thick struts we performed simulations for various pump pulse durations to assess the contribution of other nonlinear processes, as the pump pulse shortens. The fiber geometry with this particular set of parameters was selected, because it allowed in simulations to obtain the idler component at a wavelength longer than 3000 nm (that is further than the OH⁻ absorption peak of the fiber glass) and the -20 dB intensity difference relative to the pump intensity was numerically achieved for a 10 ps pump pulse with 12 nJ. Simulations for the same input energy, but with shorter pulses of 6 ps and 3 ps respectively, shown with the 10 ps pumping scenario in figure 8, demonstrated that self-phase modulation becomes a factor detrimental to the parametric conversion performance. In particular, for the 3 ps pulses, the pump signal becomes significantly broadened by self-phase modulation with a blue-shifted optical wave breaking wing already developing as well. At the same time the idler component intensity drops to -30 dB and -40 dB relative to the intensity of the broadened pump feature for the 6 ps and 3 ps long input pulses, respectively.

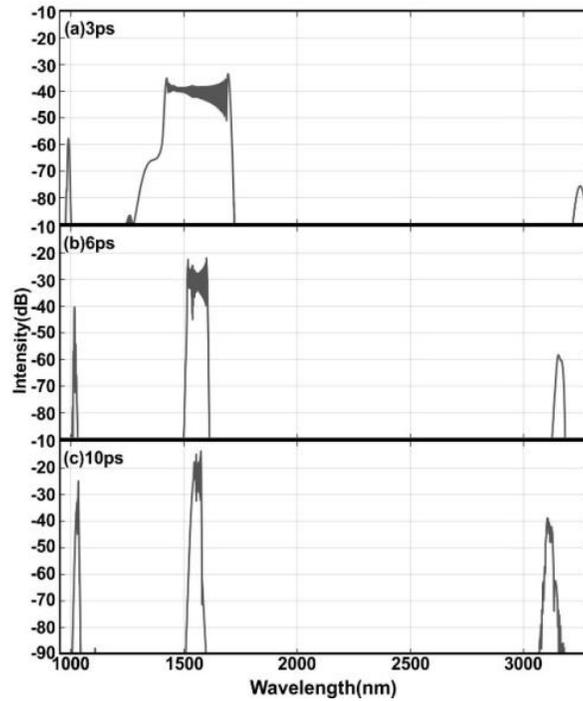

**Figure 8** Nonlinear simulations for a fiber having core diameter of 4 μm and 0.8μm thick struts, when pumped with 1560 nm having energy 12 nJ with varying pulse width at a) 3 ps, b) 6 ps and c) 10 ps. Self-phase modulation and optical wave breaking take over for shorter pulses.

## 6. Characterization of developed fibers

A test series of suspended core fibers has been developed. Geometric parameters of the developed fibers are shown in Table 1. Preforms of the fibers were prepared from lead-bismuth-gallium glass rods and tubes. Standard stack-and-draw method of PCF fabrication was used. Obtained fibers had core diameters between 3 μm and 5 μm, which was consistent with the discussed designs. Strut thicknesses obtained in technological processes were in the range of 0.1-0.3 μm, which is at the lower limit of the range assumed in the modelling. The wavelength dependence of the effective area was more favorable (weaker) for parametric conversion for structures with thin struts. Chromatic dispersion characteristics were measured in a broad wavelength range using a Mach-Zehnder interferometer and approach proposed in [30] was used to interpret the obtained interferograms. The obtained chromatic dispersion characteristics are shown in figure 9. Location of the ZDW varied in the developed fibers from about 1464 nm for a fiber with core diameter of 3.18 μm to 1705 nm for a fiber with core diameter of 5.37 μm. Generally, the dispersion profiles were obtained for a wavelength range of 1200-2000 nm with certain inconsistency resulting from limited interferogram fringe visibility in case of some fibers. This was attributed small differences in coupling of the reference light into individual fiber samples, resulting from eg. imperfect cleave. Nevertheless, the approach proposed by Hlubina et al. in [30] allowed for precise measurement of the ZDW in each of the fiber samples.

**Table 1.** Geometric parameters of fibers developed for parametric conversion under 1560 nm pump.

|   | Label | Core diameter (μm) | Strut thickness (nm) | ZDW (nm) |
|---|-------|--------------------|----------------------|----------|
| 1 | NL44C1a | 4.53 | 81 | 1625 |
| 2 | NL44C2a | 4.27 | 99 | 1585 |
| 3 | NL44C3a | 3.95 | 89 | 1551 |
| 4 | NL44C4a | 3.59 | 65 | 1525 |
| 5 | NL44C5a | 3.26 | 115 | 1484 |
| 6 | NL44C6a | 3.18 | 59 | 1464 |
| 7 | NL44B2c | 5.37 | 192 | 1704 |
| 8 | NL44B3a | 5.23 | 424 | 1693 |
| 9 | NL44B3d | 5.14 | 306 | 1690 |

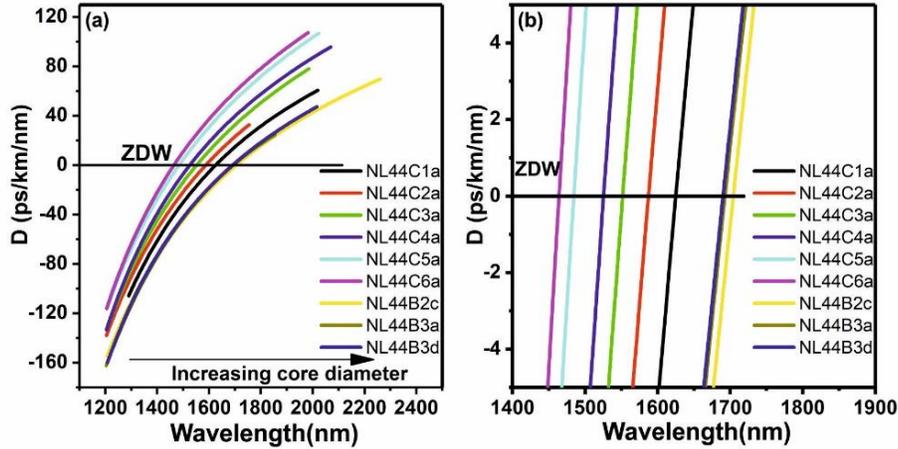

**Figure 9**. Measured chromatic dispersion characteristics of the developed fibers.

Experiments with sub-picosecond pulse pumping at a wavelength of 1560 nm were performed with fibers labelled NL44B3a (ZDW at 1693 nm), NL44C1 (1625 nm) and NL44C2a (1585 nm). Fiber samples were cleaved at a length of 20 cm, motivated by convenience of handling. The pump laser used here was similar to the one described earlier [31] In essence, this was an erbium-doped fiber-based, chirped pulse amplification laser, mode-locked with a graphene saturable absorber. The laser delivered about 600 fs long pulses, with a center wavelength of 1560 nm. Performance of the such system with nonlinear photonic crystal fibers made of a similar composition of heavy metal oxide glass, has already been investigated in context of supercontinuum generation [26]. Coupling to the fibers was obtained with a black diamond, mid-infrared aspheric lens from Thorlabs (f= 4 mm, NA = 0.56). Since the far-detuned signal intensity was expected to be difficult to detect, first the blue-shifted signal component was searched, using a compact InGaAs-based spectrometer with 900-1700 nm sensitivity range.

## 7. Experimental results and discussion

The results obtained for each of the three fibers are shown in figure 10. Fibers NL44B3a and NL44C1a, shown in figure 10 a) and b), reveal features which can be assigned to a parametric signal and this is supported by the ZDW of each of the fibers being red-shifted from the pump wavelength. In the case of the NL44C2a fiber, efficient supercontinuum generation was observed starting with very low input power, because the pump laser wavelength closely matches this fiber's ZDW. Table 2 summarizes three lowest order, even dispersion terms obtained from the dispersion profiles. The central wavelength of the pump pulses falls into normal dispersion wavelengths for all fibers and hence, values of $\beta_2$ are positive for all of them. Negative values of $\beta_4$ required for parametric gain, were obtained only for fibers NL44B3a and NL44C1a. For this reason, parametric conversion in the NL44C2a fiber cannot be expected, despite that according to our measurements, the fiber's ZDW at 1585 nm is still slightly redshifted from the pump central wavelength of 1560 nm.

**Table 2.** Three lowest order even dispersion terms, obtained from measured chromatic dispersion profiles of fibers selected for parametric conversion experiments.

| Fiber | $\beta_2$ (ps$^2$/m) | $\beta_4$ (ps$^4$/m) | $\beta_6$ (ps$^6$/m) |
|---|---|---|---|
| NL44B3a | $3.6646 \times 10^{-2}$ | $-2.6935 \times 10^{-6}$ | $3.5745 \times 10^{-9}$ |
| NL44C1a | $1.9782 \times 10^{-2}$ | $-7.9560 \times 10^{-7}$ | $-1.0368 \times 10^{-10}$ |
| NL44C2a | $8.9672 \times 10^{-3}$ | $1.4478 \times 10^{-7}$ | $-2.0491 \times 10^{-9}$ |

Spectra recorded for NL44B3a and NL44C1a fibers show self-phase modulation around the pump wavelength and distinct, blue-shifted features centered at 1205 nm and 1085 nm, respectively. These features can be assigned to either the signal product of a parametric process, or to a dispersive wave (DSW), related to a soliton excited by part of the pump pulse energy crossing the fiber's ZDW. In fact, soliton-like components are still resolvable at the edge of the spectrometer's sensitivity range between 1650 nm and 1700 nm in figure 10 a) and b). DSW locations have been calculated for the two fibers using the formula:

$$\frac{|\beta_2(\omega_s)|}{2t_s^2} = \sum_{m>2} \frac{\beta_m(\omega_s)}{m!}(\omega_d - \omega_s)^m \quad \#(4)$$

$$t_s = \frac{t_0}{2N-1} \quad \#(5)$$

where $\omega_s$ and $\omega_d$ are the soliton and DSW center angular frequencies respectively, $t_s$ and $t_0$ are the soliton and the pump pulse durations and $\beta_2$ is the group velocity dispersion at the respective frequency. N stands for input soliton order:

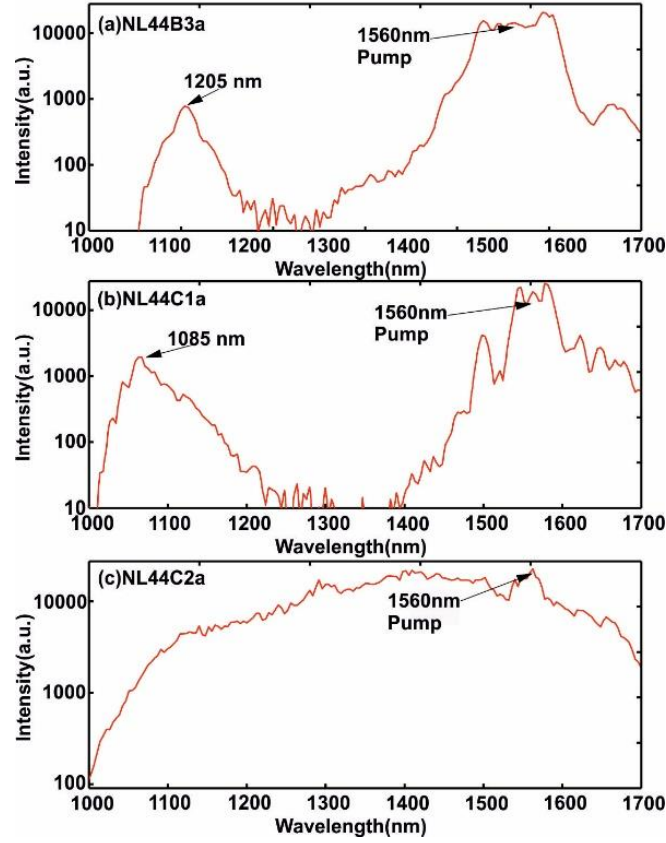

**Figure 10.** Experimentally obtained near-infrared spectra in a) NL44B3a, b) NL44C1a, c) NL44C2a fibers under 1560 nm, 600 fs pumping.

$$N = \frac{t_0^2 \gamma P_0}{|\beta_2|} \quad \#(6)$$

The parameters used in calculations corresponded to experimental conditions as follows: the peak power of the in-coupled pulse was $P_0 = 6$ kW, because of very similar core diameters, one nonlinear coefficient was assumed for both fibers, $\gamma = 500$ km$^{-1}$W$^{-1}$. Calculated DSW wavelengths are shown in figure 11 for a range of soliton central wavelengths of 1650 nm to 1800 nm. For the NL44B3a fiber the calculated DSW stays within the 1200-1220 nm range, which roughly corresponds to the spectral feature recorded for this fiber, as shown in figure 10 a). For the NL44C1a fiber, the result is very similar and the DSW remains above the wavelength 1160 nm, which does not match the 1085 nm center wavelength recorded for this fiber as shown in figure 10 b). With this, the experiment with both fiber was repeated with the InGaAs spectrometer replaced with a Fourier optical spectrum analyzer covering 1000-5600 nm wavelengths (Thorlabs OSA 205), however no significant spectral features were recorded, which was attributed to the fact that possible red-shifted idler features were below the sensitivity threshold of this device. Measurements with both fibers were repeated using again a diffraction-based optical spectrum analyzer, yet at the time of experiments the one remaining device at our disposal covered wavelengths of 1200-2400 nm range (Yokogawa AQ6375). The results are shown in figure 12, yet they do not reveal any parametric idler component at wavelengths up to 2400 nm. In both cases, Raman scattering component was recorded centered at about 1900 nm, which corresponds to the first order component of Raman scattering for the fiber glass. For the NL44C1a fiber, a weak second order Raman component can be also observed at a wavelength close to 2100 nm. The following reasons for failing to observe the parametric idler component can be considered, despite the fact of confirming a signal component in case of the NL44C1a fiber. Firstly, the fiber glass material dispersion has been well established mainly for the short-wave near-infrared wavelengths, up to 1700 nm and consequently the Sellmeier coefficients used in the linear simulations are a numerical approximation for the NIR/MIR wavelengths of interest with these fibers. This introduces uncertainty into the evaluation of effective mode area, which can be larger (thus nonlinearity would be smaller) and the fibers would require higher pulse energies to induce MI. The phase-matching condition (3) for a scenario when the pump wavelength falls into normal dispersion wavelengths of the fiber, is satisfied for very large detunings when additionally the offset between the pump wavelength and the redshifted ZDW of the fiber is large, ie. in the hundreds of nm in the near-infrared. However, in such cases the mode area mismatch would exclude any parametric conversion, eg.[18]. Moreover, in the case of fibers discussed in this work, strong OH⁻ absorption excludes the range of roughly 2850-3000 nm. The expected idler position for the NL44C1a fiber is indeed below this high-loss range, but even small uncertainty in the location of ZDW, related to eg. measurement uncertainty resulting in a different ZDW in the fiber used in the experiment would relocate the idler well into that range, while the signal position would not change significantly, based on data in figure 3 and figure 6. In such a case the idler would not be observed because of overlap with the OH⁻ attenuation band. Another possible explanation is related to the pulse durations available from the pump laser, which was at our disposal during the experiments. The laser system generated 600 fs pulses under which significant part of pulse energy was consumed by SPM before DFWM could take place. The other option on our hands was bypassing the grating-based pulse compressor, which left the output pulse at a duration of roughly 170 ps. Under this pumping however, the peak power was insufficient to observe any spectral change in the input pulses.

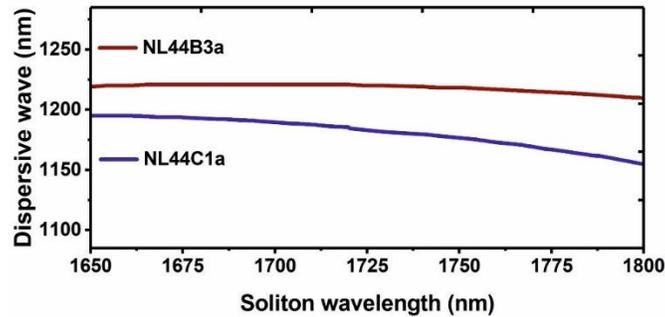

**Figure 11**. Dispersive wave locations calculated from the measured dispersion profiles for the NL44B3a and NL44C1a fibers. The minimum value shown on the Y-axis is set to the 1085 nm, corresponding to the parametric signal wavelength recorded for fiber NL44C1a.

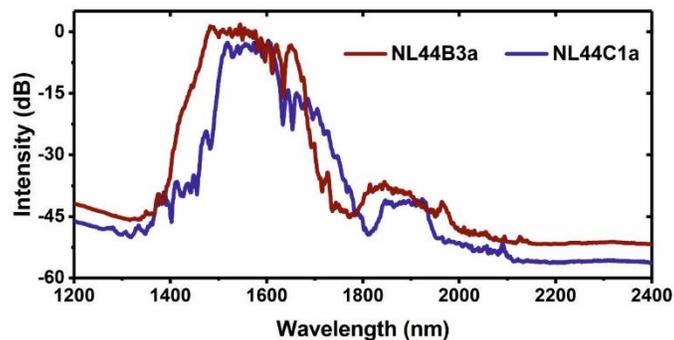

**Figure 12**. Long-wave near-infrared spectra recorded for the NL44B3a and NL44C1a fiber samples under 1560 nm, 600 fs pumping.

## 8. Conclusions

A test series of fibers for far-detuned parametric conversion, compatible with a laser operating at a standard wavelength, has been designed and developed. The primary design criterion for the fibers was obtaining dispersion profiles with ZDW redshifted from 1560 nm central wavelength of operation of erbium-doped, fiber-based CPA lasers. We discussed the shapes of obtained chromatic dispersion characteristics in context of the location of the signal and idler parametric wavelengths. GNLSE simulations of nonlinear propagation were performed for selected fiber geometries indicating parametric conversion idler components in a broad wavelength range between 3100 nm and 3500 nm at a moderate -20dB intensity of the idler relative to the pump peak. This was obtained for a reasonable range of 12-20 nJ pump pulse energies (3-10 ps durations), which can be readily obtained from different variants of erbium-doped fiber-based CPA systems.

Depending on the core size, the ZDW obtained experimentally for the physically developed fibers varied between 1464 nm and 1704 nm. The fiber line-up was shortlisted to 3 samples with ZDWs at 1585 nm, 1625 nm and 1693 nm for parametric conversion experiments. Pump was delivered with 600 fs long pulses centered at 1560 nm. Two fiber samples enabled to record blue-shifted spectral components which could be assigned to either a DSW or a DFWM signal components, which was determined by calculating DSW phase matching wavelengths. Unfortunately, no redshifted, parametric idler components were observed. The main reasons include unfavorable combination of pump pulse duration setting and available peak power and limited sensitivity of used spectrometers. The developed fibers nevertheless show promise with the parametric signal component appearing close to a wavelength typical for ytterbium-based lasers, at pumping with an erbium fiber laser. The further steps should include characterization of the refractive index dependence on the short-wave mid-infrared wavelengths, together with development of fibers with core sizes which would allow a denser sampling of the ZDW locations, and thus signal/idler locations, as well introducing a continuous-wave seed at the signal wavelength to increase intensity at the expected idler.

**Funding Information**
SONATA project UMO-2013/11/D/ST7/03156 awarded by National Science Centre in Poland; First TEAM/2016-1/1 project awarded by the Foundation for Polish Science Team Programme from the funds of European Regional Development Fund under Smart Growth Operational Programme.

**Acknowledgment**
Dariusz Pysz and Bartłomiej Siwicki from Institute of Electronic Materials Technology are acknowledged for drawing of the suspended core fibers.